\documentclass[traditabstract]{aa}

\usepackage{natbib}
\usepackage{graphicx}
\usepackage{amssymb}
\usepackage{amsmath}
\usepackage[dvipsnames]{color}

\begin{document}

\newcommand{\3}{\ss}
\newcommand{\n}{\noindent}
\newcommand{\eps}{\varepsilon}
\newcommand{\be}{\begin{equation}}
\newcommand{\ee}{\end{equation}}
\newcommand{\bl}[1]{\mbox{\boldmath$ #1 $}}
\def\ba{\begin{eqnarray}}
\def\ea{\end{eqnarray}}
\def\de{\partial}
\def\msun{M_\odot}
\def\msol{M_\odot}
\def\te{T_{\rm eff}}
\def\logg{\log g}
\def\lmix{l_{\rm mix}}
\def\lint{l_{\rm mix}^{\rm int}}
\def\latm{l_{\rm mix}^{\rm atm}}
\def\Hp{H_{\rm P }}
\def\he3{^3He}
\def\mv{M_{\rm V}}
\def\mi{M_{\rm I}}
\def\mj{M_{\rm J}}
\def\mk{M_{\rm K}}
\def\lsol{L_\odot}
\def\lbol{L_{\rm bol}}
\def\div{\nabla\cdot}
\def\grad{\nabla}
\def\rot{\nabla\times}
\def\ltsima{$\; \buildrel < \over \sim \;$}
\def\simlt{\lower.5ex\hbox{\ltsima}}
\def\gtsima{$\; \buildrel > \over \sim \;$}
\def\simgt{\lower.5ex\hbox{\gtsima}}

\newcommand{\cp}{\citep}
\newcommand{\ct}{\citet}
\newcommand{\cta}{\citetalias}

\title{A closer look at the transition between fully convective and partly radiative  low mass stars}
\titlerunning{The transition between fully  convective and partly radiative low mass stars}

\author{Isabelle Baraffe \inst{1,2}, and Gilles Chabrier \inst{2,1} }
\authorrunning{Baraffe and Chabrier}

\offprints{I. Baraffe} 
   
\institute{
University of Exeter, Physics and Astronomy, EX4 4QL Exeter, UK
(\email{i.baraffe@ex.ac.uk})
\and
\'Ecole Normale Sup\'erieure, Lyon, CRAL (UMR CNRS 5574), Universit\'e de Lyon, France
(\email{chabrier@ens-lyon.fr})
}

\date{}

\abstract{Recently, Jao et al. (2018) discovered a gap  in the mid-M dwarfs main sequence revealed by the analysis of Gaia data Release 2. They suggested the feature is linked to the onset of full convection in M dwarfs. Following the announcement of this discovery, MacDonald \& Gizis (2018) proposed an explanation  based on standard stellar evolution models. In this paper we re-examine the explanation suggested by MacDonald \& Gizis (2018).  
We confirm that nuclear burning and mixing process of $^3$He provide the best explanation for the observed feature.
We also find that a change in the energy transport from convection to radiation does not induce structural changes that could be visible. Regarding the very details of the process, however, we disagree with  MacDonald \& Gizis (2018) and propose a different explanation.}


\maketitle

\section{Introduction}
The wealth of precise all-sky data from the Gaia data Release 2 (DR2) revealed a new feature in the Herzsprung-Russell Diagram (HRD), namely a gap in the mid-M dwarfs main sequence \cp{Jao18}. The gap appears at a magnitude $M_{\rm G} \sim 10$ and colour $G_{\rm BP} - G_{\rm RP} \sim 2.3-2.5$ in the Gaia filter system. It is observed in optical and near-infrared colour-magnitude diagrams (CMDs), indicating that it is not specific to the Gaia photometry and not due to an atmospheric feature that would depend on the wavelength. \ct{Jao18} suggest the feature is linked to the onset of full convection in M dwarfs. Interestingly enough, \ct{Mayne10} was the first to suggest the existence of an observational signature for the transition between fully and partly convective structures for pre-main sequence stars and predicted that it would result in a HRD gap.  \ct{Mayne10}  explored signatures of this transition in young clusters and linked the growth of a radiative core to rapid change in effective temperature caused by changes in the dominant energy transport mechanism and ignition of hydrogen burning. Following the announcement of the HRD gap discovery by \ct{Jao18}, \ct{MacDonald18} proposed an explanation  based on standard stellar evolution models. They suggest that the observed feature is due to the complex interplay between production of $^3$He and its transport by convection.  More specifically, they predict a fast change of the luminosity in a narrow mass range  $\sim$  0.31 $\msol$ - 0.34 $\msol$  characterised by the presence of a convective core and a convective envelope, that ultimately merge. During this merging process, the central $^3$He abundance increases, causing an increase in luminosity and thus an observable feature in the luminosity function. 
In this paper, we re-examine the explanation suggested by \ct{MacDonald18} since at first sight, it is not clear why a sudden increase of the luminosity due to the merging of the core-envelope convective zones would create a dip in the luminosity function, and thus a gap in the HRD. In this analysis, we confirm that the best explanation for the observed feature is linked to the property of $^3$He nuclear production and destruction and to its mixing. We also find that a change in the energy transport from convection to radiation does not induce structural changes that could be visible. Regarding the very details of the process, however, we disagree with  \ct{MacDonald18}  and propose a different explanation. 

\section{Evolutionary Models}

Evolutionary calculations are based on the evolutionary code and input physics appropriate to the interior structures of low mass stars and substellar objects described in detail in \ct{Chabrier97} and \ct{Baraffe98}. We briefly recall below the main input physics. Models are based on the equation of state from \ct{Saumon95},  the Rosseland mean opacities of \ct{Iglesias96} and \ct{Ferguson05}. 
Convective energy transport is described by the mixing length theory, adopting a mixing length $l_{\rm mix} = 1.6 \times H_{\rm P}$, with $H_{\rm P}$ the pressure scaleheight. The onset of convective instability is determined by the Schwarzschild criterion. Outer boundary conditions are based on the atmosphere models described in \ct{Baraffe15}. Models have an initial helium abundance  $Y=0.28$, in mass fraction, and solar metallicity.
The only difference in this work compared to the recent models of \ct{Baraffe15} is the refinement of the mass grid between 0.25 $\msol$ and 0.4 $\msol$, using presently a spacing of 0.01 $\msol$ instead of 0.1 $\msol$ previously, to better describe the fully/partly convective transition. 

\subsection{Evolution with time: contraction toward the main sequence}

\begin{figure}[h!]
\vspace{-1.cm}
\includegraphics[height=12cm,width=9cm]{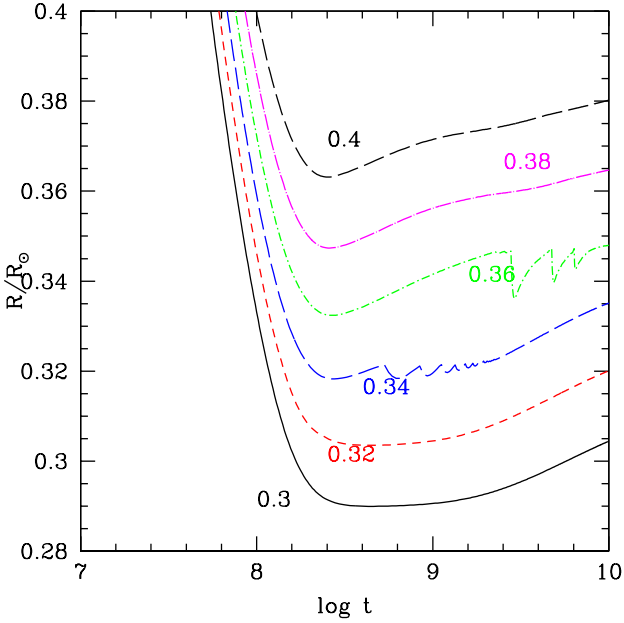}
\vspace{-2.2cm}
   \caption{Evolution of the radius with time (in yr) for low mass star models in the transition region between
   fully  and partly convective structures. Masses (in $\msol$) are indicated close to each corresponding curve.}
  \label{tr_fig}
\end{figure}

\begin{figure}[h]
\vspace{-1.5cm}
\includegraphics[height=14cm,width=9.5cm]{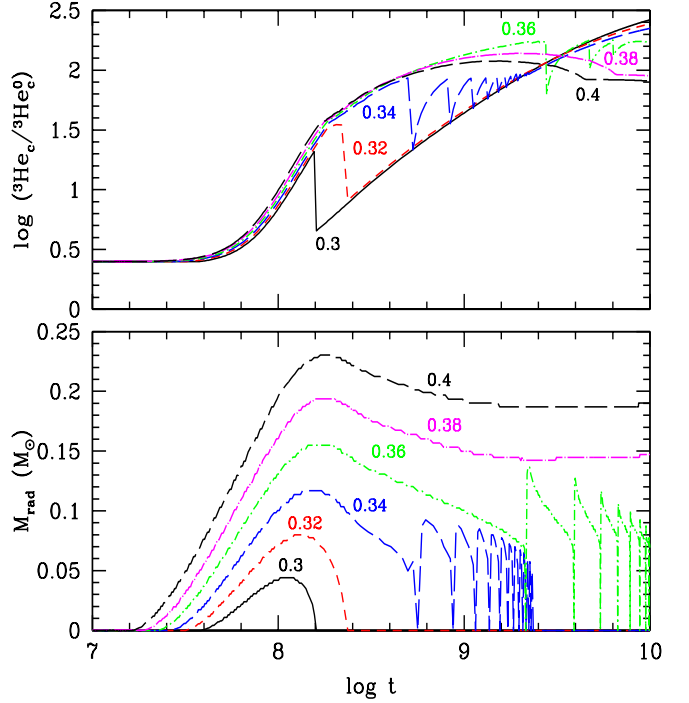}
\vspace{-2.5cm}
   \caption{Top panel: Evolution with time (in yr) of the central $\he3$ abundance (in $\log$ and normalised to the initial $\he3$ abundance $\he3_0 = 2 \, 10^{-5}$ in mass fraction). Bottom panel: evolution of the mass of the radiative core. Same masses as in Fig. \ref{tr_fig}.}
  \label{the3c_fig}
\end{figure}

Figure \ref{tr_fig} shows the evolution of the radius as a function of time in the transition region.
All models with $M \sim$ 0.28 $\msol$ start to develop a radiative core, the lower the mass the later the time the radiative core appears. As explained in \ct{Chabrier97}, the development of a radiative core stems from the decrease of the opacities after the last bump due to metals,  for central temperatures $T_{\rm c} \simgt 4 \,10^6$ K. In the mass range $\sim$ 0.28 $\msol$ -  0.35 $\msol$, the radiative core eventually disappears and the model becomes fully convective (see Fig. \ref{the3c_fig}).  Models with masses $M \simgt$ 0.33 $\msol$ also develop a convective core that will merge at some point with the convective envelope for $M \simlt 0.35 \msol$, as noticed by \ct{MacDonald18} . 

The evolution of the central abundance of $^3{\rm He}$ is particularly important in the present context. In this mass range, nuclear energy is produced by the PPI chain:

$$p+p\rightarrow d+e^++\nu_e   \eqno(1)$$
$$p+d\rightarrow ^3He\,+\,\gamma  \eqno(2)$$
$$^3He+^3He\rightarrow ^4He\,+\, 2p  \eqno(3)$$

We use the thermonuclear reaction rates from \ct{Caughlan88}, which for these specific reactions are very similar to the ones provided by another widely used compilation from \ct[][NACRE]{Angulo99}.  In the relevant temperature range (10$^6$ K - 10$^7$ K),
the rates between these two compilations differ by less than 5\%, 22\% and 7\% for reactions (1), (2) and (3) respectively.
Nuclear energy production  by reaction (3) starts to be important  for
temperatures $T \simgt 7.7 \, 10^6$K, i.e for $M \simgt 0.33 \msol$. Contribution of reaction (3) to the total nuclear energy production is what drives the development of a convective core. A simple test setting the energy production by reaction (3) to zero indeed suppresses the development of a convective core in the mass range of interest. As long as $\he3$ has not reached equilibrium, its abundance increases with temperature. Once equilibrium is reached, the abundance of $\he3$ decreases as temperature increases \cp[see e.g][]{Clayton68}.  At the onset of the convective core, the lifetime of $\he3$ against destruction by reaction (3) is $\sim 3 \,10^8$ yr (e.g for 0.4 $\msol$) to $\sim$ $10^9$ yr (e.g for 0.33 $\msol$) in the very centre and increases rapidly as a function of radius as temperature decreases toward the outer layers. As the convective core grows in mass, the average lifetime of $\he3$  against destruction in the central core is still too long, compared to the age of the model, to enable $\he3$ to reach equilibrium. 
This implies that as central temperature increases during the contraction phase toward the main sequence (typically for ages $\simlt$ 1 Gyr), the central abundance of $\he3$ increases with time (see Fig. \ref{the3c_fig}). In addition, because $\he3$ has still not reached equilibrium in the convective core (and thus its abundance increases with temperature), its abundance is maximum in the convective core and decreases as a function of radius, with a lower abundance in the convective envelope.


\begin{figure}[h]
\vspace{-1cm}
\includegraphics[height=12cm,width=9cm]{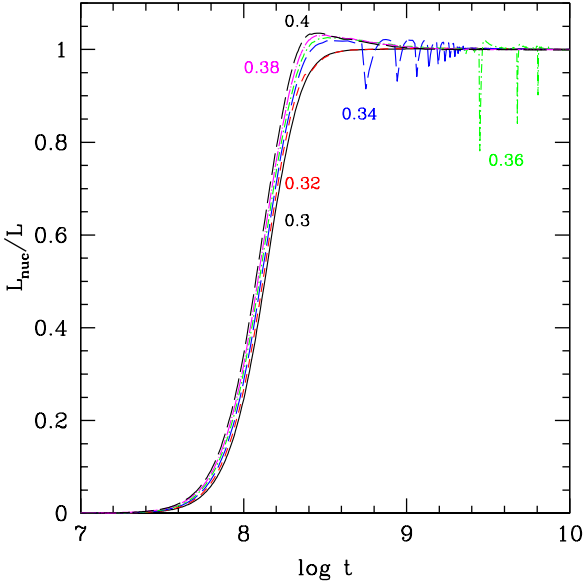}
 \vspace{-2.2cm}
   \caption{Evolution with time (in yr)  of the nuclear luminosity divided by the total luminosity. Same masses as in Fig. \ref{tr_fig}.}
  \label{tlnuc_fig}
\end{figure}

For masses $M \simgt 0.34 \msol$, reaction (3) provides a significant contribution to the total nuclear energy production by the PPI chain. The continuous increase with time of the central abundance of $\he3$ results in an overproduction of nuclear energy in the central regions for masses $M \simgt 0.34 \msol$.  Consequently, the nuclear luminosity $L_{\rm nuc}$ exceeds the total luminosity of the star $L$ (see Fig.  \ref{tlnuc_fig}) as the model approaches the beginning of the main sequence  (that can be defined by thermal equilibrium with $L_{\rm nuc} \sim L$). This yields a noticeable expansion (i.e increase of the radius) of the models  at ages $\log t \sim 8.4 - 9$ compared to their lower mass, fully convective, counterparts (see Fig.  \ref{tr_fig}). 

For models with $M \simlt 0.33  \msol$, which become fully convective when reaching the main sequence, the contribution of equation (3) to the nuclear energy production remains small. When the radiative core disappears, the strong drop of central abundance of $^3{\rm He}$, due to the mixing of envelope material that contains lower abundance of $^3{\rm He}$ compared to the central regions, has no particular effect on the total nuclear luminosity, as seen in  Fig.  \ref{tlnuc_fig}. 

The intermediate cases ($M \sim 0.34  \msol - 0.36 \msol$) show more complex behaviours with burst events due to episodic merging of the convective core and the envelope. 
This feature was initially discovered by \ct{Vansaders12} who report the existence of a new instability near the fully convective boundary driven by $^3{\rm He}$. \ct{Vansaders12} have analysed in detail this instability named the "convective kissing instability". We confirm its existence using different stellar evolution code and input physics.
At each merging event, the central abundance of $\he3$ suddenly drops, since central regions with higher abundance of $\he3$ mix with envelope material containing less $\he3$. This is where our results differ from \ct{MacDonald18}.  The strong drop of  central   $^3{\rm He}$  results in a drop of the nuclear energy production, as can be seen in the drop of $L_{\rm nuc}$ in Fig. \ref{tlnuc_fig} for the displayed 0.34 $\msol$ and 0.36 $\msol$ models. The effect is an overall contraction, impacting the radius evolution.  Note that the Kelvin-Helmholtz timescale, $\tau_{\rm KH} \sim G M^2/(RL)$, is of the order of a few 10$^8$ years for these low mass stars approaching the main sequence, indicating that change of conditions in the central region has effectively the time to induce the structural change, in terms of radius and thus luminosity, that we describe.

\subsection{Mass-radius and mass-luminosity relationships}

\begin{figure}[h]
\vspace{-1cm}
\includegraphics[height=13cm,width=9cm]{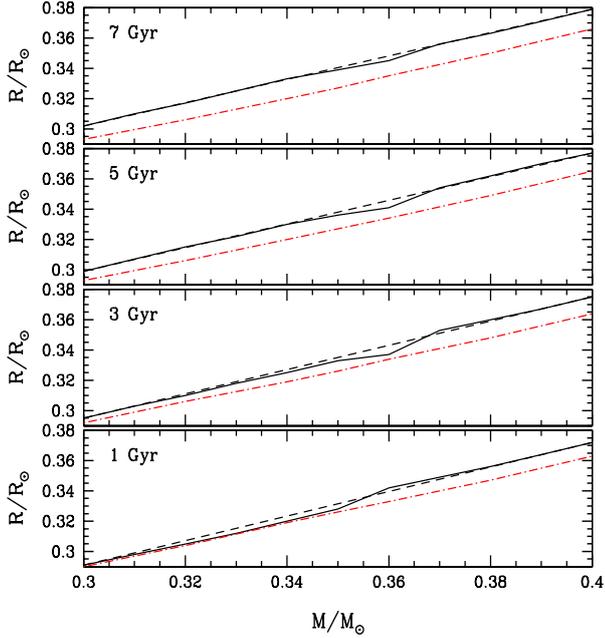}
\vspace{-2.2cm}
	\caption{Mass-radius relationships at different ages in the fully/partly convective transition region (solid line). Note that the mass grid uses a spacing  of 0.01 $\msol$. To highlight changes in the slope, the dashed curve is a straight line connecting values of the radius between 0.3 $\msol$ and 0.4 $\msol$. The dash-dotted (red) curves correspond to the test case where $\he3$ is forced to reach equilibrium (see text \S 2.2).}
  \label{mr_fig}
\end{figure}

\begin{figure}[h]
\vspace{-1cm}
\includegraphics[height=13cm,width=9cm]{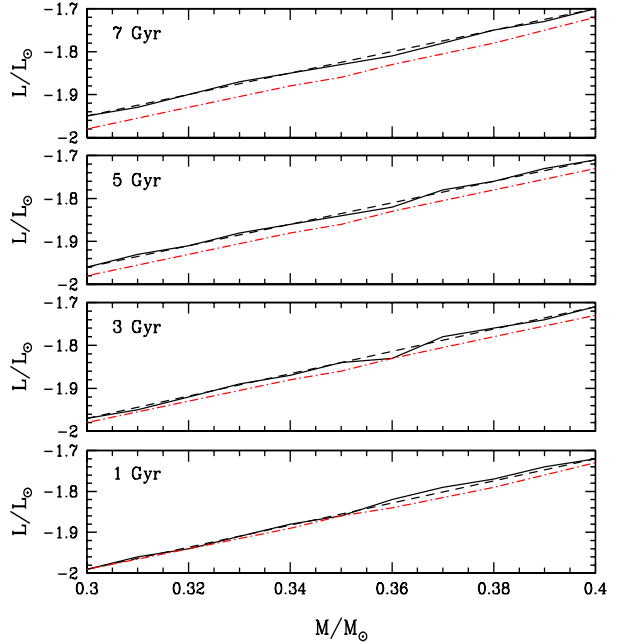}
\vspace{-2.5cm}
   \caption{Mass-luminosity relationships at different ages in the fully/partly convective transition region (solid line). The other curves have the same meaning as in Fig. \ref{mr_fig}.}
  \label{ml_fig}
\end{figure}

Understanding the evolution with models of different masses helps grasping the properties of the mass-radius and mass-luminosity relationships discussed in this section. We suggest that two features shape these relationships.  The first one (feature 1) is the expansion of models with  $M \simgt 0.34 \msol$ due to the fast increase of central $\he3$ abundance, causing a fast increase of nuclear energy production. This evolution is different from the one followed by  fully convective models with $M \simlt 0.33 \msol$, which evolve at essentially constant radius during the first Gyr of evolution. The second feature (feature 2) is the significant drop in radius during the episodic event of convective core/envelope merging, which only concerns the narrow mass range $M \sim 0.34  \msol - 0.36 \msol$. Inspection of the mass radius (Fig. \ref{mr_fig}) and mass-luminosity (Fig. \ref{ml_fig}) relationships at different ages reveals these two features. Feature 1 is most visible for ages $\simlt 3$ Gyr, with the $M-R$ and $M-L$ relationships  lying slightly above the straight dashed line (added as a guide for the eye) for $M >  0.36 \msol$ and lying below it for $M < 0.35 \msol$. This is the signature of the expansion that only proceeds above the fully/partly convective transition. Feature 2 could be seen at all ages between $\sim$ 1-7 Gyr , contributing to a dip in the $M-R$ and $M-L$ relationships, but in a very narrow mass range. 

In order to verify the key role played by $\he3$ destruction/production  in the present context, we perform a test in which $\he3$ is forced to reach equilibrium. This is achieved by multiplying the reaction rate of reaction (3) by a factor 10 (but the reaction rate used for the calculation of the nuclear energy production is unmodified). This forces $\he3$ to reach equilibrium during the contraction phase toward the main sequence, with the central abundance of $\he3$ rapidly reaching a maximum. The approach to the main sequence proceeds without the strong expansion reported for $M \simgt 0.35 \msol$. In this test, the transition between fully/partly convective models takes place at $M \sim$ 0.34 $\msol$.  Convective cores do not develop because of the limitation of $\he3$ production. There is thus no episodic events of convective core/envelope merging. The evolution in the transition region is smooth. No change of slope in the $M-R$ and $M-L$ relationships is observed in this test case (see dash-dotted curves in Fig. \ref{mr_fig} and Fig. \ref{ml_fig}). This confirms the key role played by $\he3$ in the transition region. It also confirms that {\it the change of energy transport from convective to radiative transport in the central regions does not induce any visible structural change}. 

\subsection{Colour-Magnitude diagram}

\begin{figure}[h!]
\vspace{-1.5cm}
\includegraphics[height=12cm,width=9cm]{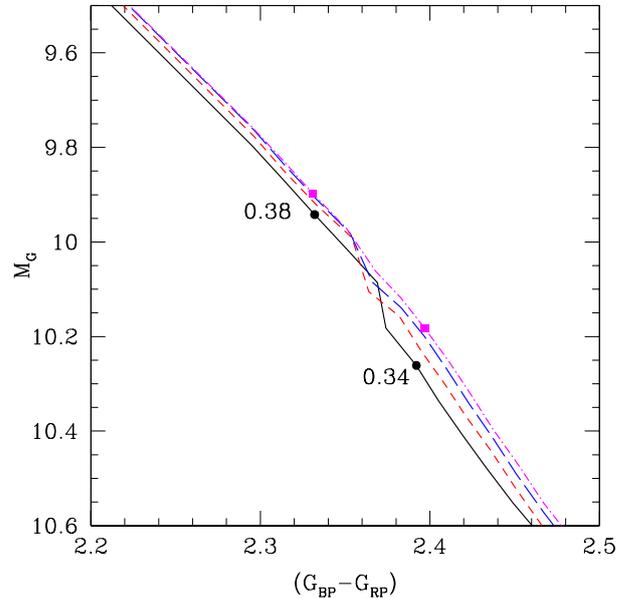}
\vspace{-2.5cm}
   \caption{Colour-Magnitude diagram in the Gaia filter system for various isochrones. Black solid line: 1 Gyr; Red dash: 3 Gyr; Blue long dash: 5 Gyr; Magenta dash-dot: 7 Gyr. The symbols on the 1-Gyr and 7-Gyr curves indicate the position of models with 0.34 $\msol$ and 0.38 $\msol$, respectively.}
  \label{maggaia_fig}
\end{figure}

The features described in the previous section are visible in the colour-magnitude diagram using the same Gaia filters as in \ct{Jao18}. The changes of slopes and the drop in magnitude for isochrones between 1-7 Gyr take place at magnitudes $M_{\rm G} \sim 10 -10.2$ and colours  
 $G_{\rm BP} - G_{\rm RP} \sim 2.3 - 2.4$ (see Fig. \ref{maggaia_fig}), which nicely correspond to the location of the gap reported by \ct{Jao18}.

\section{Discussion and conclusion}

Confirmation that the two evolutionary features described in the previous section are the genuine explanation for the gap observed in the Gaia data requires the construction of synthetic colour-magnitude diagrams. This will allow testing their observational and statistical relevance. This is beyond the scope of present analysis, which primarily aims at clarifying the explanation provided by \ct{MacDonald18}.  Given the very narrow mass range of  models experiencing feature 2, i.e the convective core/envelope merging, this may not be the most relevant feature in terms of statistics. It could, however, further contribute to the change of slope due to feature 1 in the $M-R$ and $M-L$ relationships predicted by models above $M > 0.36 \msol$ and below $M < 0.34 \msol$ 
The evolutionary features we find  should be robust against changes in the input physics and variations of the metallicity around solar values. We verified this by running models in the same mass range adopting the input physics of \ct{Baraffe98},  which use different atmosphere models for the outer boundary conditions and allow a comparison of results between solar [M/H]=0 and subsolar  [M/H]=-0.5 metallicities. The same qualitative evolutionary features are found with the \ct{Baraffe98} input physics for both metallicities.
The exact location of the transition, in terms of mass and luminosity, between fully/partly convective stars, as well as the strength and number of convective core/envelope merging events will, however, depend on the input physics, the metallicity and likely also on different numerical treatments of mixing in stellar evolution codes.  We also find quantitative differences in the properties of the convective core/envelope merging events (e.g time when the first event happens) depending on the choice of the numerical timestep and the grid resolution. \ct {Vansaders12} present a  detailed analysis of this instability, which is not the main focus of our study, and the quantitative differences that may result from different models. 
 
Our interpretation differs from the one provided by \ct{MacDonald18}. We find that the merging of the convective core with the envelope results in a {\it decrease} of $\he3$ abundance in the centre and thus a decrease in luminosity. 
 \ct{MacDonald18} report the opposite because they find that $\he3$ reaches quasi-equilibrium in the convective core in the initial main sequence evolution and that the abundance of $\he3$ is larger above the convective core.
 Estimate of the {\it average} lifetime of $\he3$ against destruction by reaction (3) in the convective core and the test we performed by forcing equilibrium suggest that the latter has no time to be reached during the approach to the main sequence. We thus do not agree with their interpretation.
 
The discovery of the HRD gap by \ct{Jao18} could have an impact in the field of cataclysmic variable (CV) systems (Knigge, priv. com).
The complex behaviour of the mass-radius relationship that standard evolutionary models predict at the transition could affect the evolution of low mass donors in CVs. Could this explain the famous CV period gap, namely the dearth of systems in the period range 2hr - 3hr \cp{King88, Knigge11}? The most popular explanation is the disruption of magnetic breaking, the dominant angular momentum loss mechanism for these systems above the period gap, once the low mass donor becomes fully convective. Although issues exist with this standard scenario \cp[see discussion in][]{Knigge11}, a better explanation has not been found yet.   A small drop in radius as the CV donor becomes fully convective could provide an alternative explanation to the disruption of magnetic breaking, as it would cause detachment of the system and interruption of the mass transfer (and thus a period gap). Unfortunately, though the idea is extremely attractive, the central conditions of low mass donors in CVs when they become fully convective are different from the ones of their low mass star counterparts which evolve at constant mass. Standard sequences following the evolution of low mass donors above the period gap \cp[see e.g][]{Baraffe00} show that when the donor becomes almost fully convective,  at a mass $M \sim 0.2 \msol$ and period $P \sim$ 3hr, the central temperature is too low ($T_{\rm c} \simlt 6 \, 10^6$ K) for reaction (3) to be efficient and to provide the same mechanism as found for low mass stars. The transition to a fully convective structure is thus smooth with no obvious structural change that could induce a drop in radius for CV donors.
\ct{Vansaders12} also analysed the impact of the convective envelope/core merging events (their so-called convective kissing instability) on the evolution of CV secondaries. They suggest that the instability could play a role on CV secondaries close to the period gap, but only for extremely low mass transfers. Such low mass transfers seem rather unrealistic according to the  exhaustive study of \ct{Knigge11}. In addition, their scenario will struggle to produce the observed gap width (between 2hr and 3hr). Indeed, with very low mass transfers, the instability kicks in at a typical mass of 0.35 $\msun$ \cp[see Fig. 3 of][]{Vansaders12}, 
which can provide the upper edge of a gap at 3hr if the system detaches. It seems however extremely difficult to obtain the lower edge of the gap (once the secondary shrinks back to its thermal equilibrium configuration and mass transfer resumes) at a period of 2hr with such a high mass.

Our explanations for the HRD gap still need to be confirmed with the generation of statistically significant synthetic colour-magnitude diagrams. If they are confirmed, the observed signature of this transition provides excellent diagnostics of the central conditions of  low mass star evolutionary models, since the present mechanisms highly depend on the central temperature and the evolution of both the convective core and the envelope. Rotational and magnetic properties of objects on either side of the HRD gap might also be affected, helping us understanding
 low mass star properties when they become fully convective. Just a small gap in a colour-magnitude diagram could thus provide a deep insight into the interior structure of low mass stars.

 \section{Acknowledgements}
We thank Christian Knigge, Russel White and Wei-Chun Jao for very useful discussions. This work is partly supported by the consolidated STFC grant ST/R000395/1 and the ERC grant No. 787361-COBOM.

\bibliographystyle{aa.bst}
\bibliography{references}

\begin{thebibliography}{16}
\expandafter\ifx\csname natexlab\endcsname\relax\def\natexlab#1{#1}\fi

\bibitem[{{Angulo} {et~al.}(1999){Angulo}, {Arnould}, {Rayet}, {Descouvemont},
  {Baye}, {Leclercq-Willain}, {Coc}, {Barhoumi}, {Aguer}, {Rolfs}, {Kunz},
  {Hammer}, {Mayer}, {Paradellis}, {Kossionides}, {Chronidou}, {Spyrou},
  {degl'Innocenti}, {Fiorentini}, {Ricci}, {Zavatarelli}, {Providencia},
  {Wolters}, {Soares}, {Grama}, {Rahighi}, {Shotter}, \& {Lamehi
  Rachti}}]{Angulo99}
{Angulo}, C., {Arnould}, M., {Rayet}, M., {et~al.} 1999, Nuclear Physics A,
  656, 3

\bibitem[{{Baraffe} {et~al.}(1998){Baraffe}, {Chabrier}, {Allard}, \&
  {Hauschildt}}]{Baraffe98}
{Baraffe}, I., {Chabrier}, G., {Allard}, F., \& {Hauschildt}, P.~H. 1998, \aap,
  337, 403

\bibitem[{{Baraffe} {et~al.}(2015){Baraffe}, {Homeier}, {Allard}, \&
  {Chabrier}}]{Baraffe15}
{Baraffe}, I., {Homeier}, D., {Allard}, F., \& {Chabrier}, G. 2015, \aap, 577,
  A42

\bibitem[{{Baraffe} \& {Kolb}(2000)}]{Baraffe00}
{Baraffe}, I. \& {Kolb}, U. 2000, \mnras, 318, 354

\bibitem[{{Caughlan} \& {Fowler}(1988)}]{Caughlan88}
{Caughlan}, G.~R. \& {Fowler}, W.~A. 1988, Atomic Data and Nuclear Data Tables,
  40, 283

\bibitem[{{Chabrier} \& {Baraffe}(1997)}]{Chabrier97}
{Chabrier}, G. \& {Baraffe}, I. 1997, \aap, 327, 1039

\bibitem[{{Clayton}(1968)}]{Clayton68}
{Clayton}, D.~D. 1968, {Principles of stellar evolution and nucleosynthesis}

\bibitem[{{Ferguson} {et~al.}(2005){Ferguson}, {Alexander}, {Allard}, {Barman},
  {Bodnarik}, {Hauschildt}, {Heffner-Wong}, \& {Tamanai}}]{Ferguson05}
{Ferguson}, J.~W., {Alexander}, D.~R., {Allard}, F., {et~al.} 2005, \apj, 623,
  585

\bibitem[{{Iglesias} \& {Rogers}(1996)}]{Iglesias96}
{Iglesias}, C.~A. \& {Rogers}, F.~J. 1996, \apj, 464, 943

\bibitem[{{Jao} {et~al.}(2018){Jao}, {Henry}, {Gies}, \& {Hambly}}]{Jao18}
{Jao}, W.-C., {Henry}, T.~J., {Gies}, D.~R., \& {Hambly}, N.~C. 2018, \apjl,
  861, L11

\bibitem[{{King}(1988)}]{King88}
{King}, A.~R. 1988, \qjras, 29, 1

\bibitem[{{Knigge} {et~al.}(2011){Knigge}, {Baraffe}, \&
  {Patterson}}]{Knigge11}
{Knigge}, C., {Baraffe}, I., \& {Patterson}, J. 2011, \apjs, 194, 28

\bibitem[{{MacDonald} \& {Gizis}(2018)}]{MacDonald18}
{MacDonald}, J. \& {Gizis}, J. 2018, \mnras

\bibitem[{{Mayne}(2010)}]{Mayne10}
{Mayne}, N.~J. 2010, \mnras, 408, 1409

\bibitem[{{Saumon} {et~al.}(1995){Saumon}, {Chabrier}, \& {van
  Horn}}]{Saumon95}
{Saumon}, D., {Chabrier}, G., \& {van Horn}, H.~M. 1995, \apjs, 99, 713

\bibitem[{{van Saders} \& {Pinsonneault}(2012)}]{Vansaders12}
{van Saders}, J.~L. \& {Pinsonneault}, M.~H. 2012, \apj, 751, 98

\end{thebibliography}

\end{document}